\documentclass[
    ,final            
  ]
  {aipproc}

\layoutstyle{6x9}

\begin{document}

\title{Electroweak Constraints on Effective Theories  }


\author{Zhenyu Han}{
  address={Department of Physics, University of California, Davis, CA 95616, USA}
}

\begin{abstract}
We discuss electroweak constraints on TeV scale extensions of the
standard model. To obtain model-independent results, effective
theory approach is adopted. Constraints are given on arbitrary
linear combinations of a set of dimension-6 operators that respect
the SM gauge symmetry, as well as CP, lepton and baryon number
conservation. Applications of the results are also discussed.
\end{abstract}
\maketitle


\section{Introduction}

Assuming insignificant fine-tuning to the Higgs mass, we expect
extensions of the standard model (SM) to appear at the TeV scale.
Many TeV scale models have been extensively studied, such as
supersymmetry, technicolor, extra-dimension and little Higgs models.
On the other hand, the results from electroweak precision tests
(EWPTs) remarkably agree with the SM predictions, and therefore
tightly constrain many of the TeV scale models.

To efficiently obtain electroweak constraints, an effective theory
approach \cite{Buchmuller, GW, Han:2004az,Han:2005pr}
is desirable. In this approach, one first integrates out
all new heavy states and obtains effective operators involving only
fields of the SM. From these operators, one calculates
deviations from the SM and compares them with experimental data.
Electroweak constraints are then obtained on the operator
coefficients. Once this step is done, one can constrain any model
just by calculating the coefficients of new effective operators.
Here we present a study on bounds on arbitrary linear combinations
of dimension-6 operators that could be relevant to TeV scale
physics.

\section{Operators}
We assume that just above the electroweak symmetry breaking scale,
the effective theory is that of the SM with one Higgs doublet. A
complete set of 80 independent dimension-6 operators consistent with
the $SU(3)\times  SU(2)_L \times U(1)_Y$ gauge symmetry, baryon and
lepton number conservation has been presented in
Ref.~\cite{Buchmuller}.

We are interested in constraining models of new physics pertinent to
the electroweak symmetry breaking. Processes that contribute to
flavor or CP violation have to be suppressed by  scales much higher
than the electroweak scale. Thus we impose CP conservation and
$U(3)^5$ flavor symmetry on our operators. A different $U(3)$ acts
on the  left-handed quarks and leptons as well as on the right-handed
quarks and leptons. In Ref.~\cite{Han:2005pr}, the $U(3)^5$
symmetry is relaxed to $[U(2)\times U(1)]^5$, where the third
generation is treated differently from the first two.

We also exclude operators that are not or poorly constrained by
current experiments. The remaining 21 operators are the focus of our
work and listed below. The notation is standard: $q$ and $l$
represent the three families of the left-handed quark and lepton
fields, respectively. The right handed fields are labeled $u$, $d$,
and $e$. We omit the family index which is always summed over due to
the flavor $U(3)^5$ symmetry.

The operators that contain only the gauge bosons and Higgs doublets
are
\begin{equation}
\label{eq:owbh}
 O_{W\!B}=(h^\dagger \sigma^a h) W^a_{\mu \nu} B^{\mu \nu}, \  \  \   O_h = | h^\dagger D_\mu h|^2,
\end{equation}
where $W^a_{\mu \nu}$ is the $SU(2)$ field strength, $B_{\mu \nu}$
the hypercharge field strength, and $h$ represents the Higgs
doublet. There are 11 four-fermion operators. These are
\begin{eqnarray} &&
  O_{ll}^s=\frac{1}{2} (\overline{l} \gamma^\mu l) (\overline{l} \gamma_\mu l), \ \ \
  O_{ll}^t=\frac{1}{2} (\overline{l} \gamma^\mu \sigma^a l) (\overline{l} \gamma_\mu \sigma^a l),
      \label{eq:oll} \\ &&
  O_{lq}^s= (\overline{l} \gamma^\mu l) (\overline{q} \gamma_\mu q), \ \ \
  O_{lq}^t= (\overline{l} \gamma^\mu \sigma^a l) (\overline{q} \gamma_\mu \sigma^a q),
     \label{eq:olq} \\ &&
  O_{le}= (\overline{l} \gamma^\mu l) (\overline{e} \gamma_\mu e),  \ \ \
  O_{qe}=(\overline{q} \gamma^\mu q) (\overline{e} \gamma_\mu e),
     \label{eq:olqe}  \\ &&
  O_{lu}= (\overline{l} \gamma^\mu l) (\overline{u} \gamma_\mu u),  \ \ \
  O_{ld}= (\overline{l} \gamma^\mu l) (\overline{d} \gamma_\mu d),
      \label{eq:olud} \\ &&
  O_{ee}=\frac{1}{2} (\overline{e} \gamma^\mu e) (\overline{e} \gamma_\mu e), \ \ \
  O_{eu}=(\overline{e} \gamma^\mu e) (\overline{u} \gamma_\mu u),  \ \ \
  O_{ed}=(\overline{e} \gamma^\mu e) (\overline{d} \gamma_\mu d).
      \label{eq:oeeud}
  \end{eqnarray}
   There are 7 operators containing
  2 fermions that alter the couplings of fermions to the gauge bosons
 \begin{eqnarray} &&
   O_{hl}^s = i (h^\dagger D^\mu h)(\overline{l} \gamma_\mu l) + {\rm h.c.}, \ \ \
   O_{hl}^t = i (h^\dagger \sigma^a D^\mu h)(\overline{l} \gamma_\mu \sigma^a l)+ {\rm h.c.},
     \label{eq:ohl} \\ &&
   O_{hq}^s = i (h^\dagger D^\mu h)(\overline{q} \gamma_\mu q)+ {\rm h.c.}, \ \ \
   O_{hq}^t = i (h^\dagger \sigma^a D^\mu h)(\overline{q} \gamma_\mu \sigma^a q)+ {\rm h.c.},
    \label{eq:ohq} \\ &&
      O_{hu} = i (h^\dagger D^\mu h)(\overline{u} \gamma_\mu u)+ {\rm h.c.}, \ \ \
   O_{hd} = i (h^\dagger D^\mu h)(\overline{d} \gamma_\mu d)+ {\rm h.c.},
           \label{eq:Ohud}  \\ &&
   O_{he} = i (h^\dagger D^\mu h)(\overline{e} \gamma_\mu e)+ {\rm h.c.}\,.
     \label{eq:ohe}
\end{eqnarray}
Finally, there is an operator that  modifies the triple gauge  boson
interactions
 \begin{equation}
 \label{eq:ow}
  O_W = \epsilon^{abc} \, W^{a \nu}_{\mu} W^{b\lambda}_{\nu} W^{c \mu}_{\lambda}.
\end{equation}
Eqs.~(\ref{eq:owbh}) through (\ref{eq:ow}) define our basis of the
21 operators. Adding these operators to the SM, we have the
effective Lagrangian as
 \begin{eqnarray}
 \label{Lagrangian}
   {\mathcal L}&=& {\mathcal L}_{SM}+\sum_i a_i  O_i\\
    &=& {\mathcal L}_{SM} + a_{W\!B} \, O_{W\!B} + a_h \, O_h + \ldots + a_W\,   O_W,
 \end{eqnarray}
where we have denoted the coefficients $a_i$ using the same indices
as the corresponding operators.
\section{Constraints}
We include in our analysis all relevant electroweak precision
observables (EWPOs). The three most precisely measured ones,
$\alpha$, $G_F$, and $M_Z$ are taken to be the input parameters,
from which the SM gauge couplings and the Higgs vev are inferred.
The other observables include the $W$ boson mass, observables from
atomic parity violation, deep inelastic scattering and $Z$-pole
experiments, and fermion and $W$ boson pair production data from LEP
2.

Starting from the Lagrangian (\ref{Lagrangian}), we calculate
deviations from the SM as functions of the coefficients $a_i$. Due
to the excellent agreement between the experiments and the SM, we
only need to work to linear order in $a_i$. For a given observable
$X$, we have
\begin{equation}
\label{prediction}
   X_{th}=X_{SM}+\sum_i a_i X_i,
\end{equation}
where $X_{th}$ is the prediction in the presence of additional
operators, $X_{SM}$ is the SM prediction, which is well known, and
$\sum_ia_iX_i$ are corrections from our new operators.

We then compare the theoretical predictions with the experimental
values $X_{exp}$, and calculate the total $\chi^2$ distribution. For
non-correlated measurements,
\begin{equation}
\label{eq:chi2uncorr}
  \chi^2(a_i)=\sum_X\frac{(X_{th}(a_i)-X_{exp})^2}{\sigma_X^2},
\end{equation}
where $X_{exp}$ is the experimental value for observable $X$ and
$\sigma_X$ is the total error both experimental and theoretical.

Since that each $X_{th}$ is linear in $a_i$, the total $\chi^2$ is
quadratic:
\begin{equation}
\label{chi2}
  \chi^2= \chi^2_{SM} + a_i \hat{v_i} +
          a_i {\mathcal M}_{ij} a_j.
\end{equation}
The numerical values for the matrix ${\mathcal M}_{ij}$ and the
vector $\hat{v_i}$ are our main results. It can be obtained from the
author. Given the $\chi^2$, what is remaining to be done to
constrain a model is only calculating the operator coefficients
$a_i$.

\section{Applications}

Our analysis is a systematic generalization of other
model-independent analyses of electroweak constraints. For example,
the operator $O_{WB}$ and $O_{h}$ correspond to the oblique $S$ and
$T$ parameters \cite{STU}. It is straightforward to reproduce the
$S$ and $T$ fit in Ref.~\cite{erler+langacker} using the 2 by 2
submatrix of ${\mathcal M}$ and the first two components of the
vector $\hat{v_i}$.

Our results can be easily used to constrain many models with
also non-oblique corrections. As an illustration, we apply the
results to the littlest Higgs model \cite{Arkani-Hamed:2002qy}. More
examples can be found in
Ref.~\cite{Han:2004az,Han:2005pr,Han:2005dz}.

In the littlest Higgs model, there exist new heavy gauge bosons,
quarks and scalars. After integrating out the new particles, we
obtain the following operator coefficients:

\begin{eqnarray}
a_h&=&-\frac{5(c'^2-s'^2)^2}{2 F^2}+\frac{2\lambda^2}{M_\phi^4},\nonumber\\
a_{hq}^t&=&a_{hl}^t=-\frac{(c^2-s^2)c^2}{2 F^2},\nonumber\\
a_{hf}^s&=&\frac{5s'c'(c'^2-s'^2)}{F^2}\left(Y_2^f\frac{s'}{c'}-Y_1^f\frac{c'}{s'}\right),\nonumber\\
a_{lq}^t&=&a_{ll}^t=-\frac{c^4}{F^2},\nonumber\\
a_{ff'}^s&=&-\frac{20s'^2c'^2}{F^2}\left(Y_2^f\frac{s'}{c'}-Y_1^f\frac{c'}{s'}\right)
\left(Y_2^{f'}\frac{s'}{c'}-Y_1^{f'}\frac{c'}{s'}\right).\label{littlestops}
\end{eqnarray}

The details of the littlest Higgs model and how to obtain
Eqs.~(\ref{littlestops}) are described elsewhere
\cite{Arkani-Hamed:2002qy,Han:2005dz}. We emphasize here that one
can obtain bounds on this model by simply substituting in
Eq.~(\ref{chi2}) the above coefficients , without detailed knowledge
of EWPTs. It turns out that there exist significant parameter space
where the bounds on the heavy particles' masses are mild enough to
solve the fine-tuning problem, as long as the fermions are charged
equally under the two $U(1)$ gauge groups in the model.

\section{Conclusion}
We have identified a set of dimension-6 operators that are relevant
to TeV scale extensions of the SM and can be tightly constrained by
EWPTs. We have obtained constraints on arbitrary linear combinations
of the operators, using all precisely measured EWPOs. The results
can be easily applied to constrain many TeV scale models, especially
those with heavy particles contributing to EWPOs at tree level.

\begin{theacknowledgments}
It is a pleasure to thank W. Skiba for collaboration. The author is
supported in part by the US Department of Energy grant No.
DE-FG03-91ER40674, and by the UC Davis HEFTI program.
\end{theacknowledgments}



\bibliographystyle{aipproc}   



\end{document}